\documentclass[10pt,conference]{IEEEtran}
\usepackage{cite}
\usepackage{amsmath,amssymb,amsfonts}
\usepackage{algorithmic}
\usepackage{graphicx}
\usepackage{textcomp}
\usepackage{xcolor}
\usepackage[hyphens]{url}
\usepackage{fancyhdr}
\usepackage[bookmarks=true,breaklinks=true,letterpaper=true,colorlinks,citecolor=blue,linkcolor=blue,urlcolor=blue]{hyperref}
\usepackage{multirow}
\usepackage{subfigure}
\usepackage{bbding}
\usepackage{mathtools}
\usepackage{makecell}
\usepackage{pifont}
\usepackage{siunitx}

\newcommand{\yaofix}[1]{\textcolor{black}{#1}} 

\newcommand{\yao}[1]{\textcolor{black}{#1}} 
\newcommand{\ming}[1]{\textcolor{black}{#1}}

\newcommand{\qijunre}[1]{\textcolor{black}{#1}}
\newcommand{\mm}[1]{\textcolor{black}{#1}}
\newcommand{\mmreb}[1]{\textcolor{black}{#1}}
\newcommand{\yaoreb}[1]{\textcolor{black}{#1}}
\newcommand{\qjreb}[1]{\textcolor{black}{#1}}

\newcommand{\hpcayear}{2025}

\title{Integrating Prefetcher Selection with Dynamic Request Allocation Improves Prefetching Efficiency\vspace{-.1in}}

\DeclareRobustCommand*{\IEEEauthorrefmark}[1]{%
\raisebox{0pt}[0pt][0pt]{\textsuperscript{\footnotesize\ensuremath{#1}}}}

\def\hpcacameraready{}
\newcommand\hpcaauthors{Mengming Li\IEEEauthorrefmark{1\dagger}, Qijun Zhang\IEEEauthorrefmark{1\dagger}, Yongqing Ren\IEEEauthorrefmark{2*}, Zhiyao Xie\IEEEauthorrefmark{1*}}
\newcommand\hpcaaffiliation{\IEEEauthorrefmark{1}Hong Kong University of Science and Technology, \IEEEauthorrefmark{2}Intel}
\newcommand\hpcaemail{\{mengming.li, qzhangcs\}@connect.ust.hk, yongqing.ren@intel.com, eezhiyao@ust.hk\\\vspace*{-.9cm}}

\author{
  \ifdefined\hpcacameraready
    \IEEEauthorblockN{\hpcaauthors{}}
      \IEEEauthorblockA{
        \hpcaaffiliation{} \\
        \hpcaemail{}
      }
  \else
    \IEEEauthorblockN{\normalsize{HPCA \hpcayear{} Submission
      \textbf{\#\hpcasubmissionnumber{}}} \\
      \IEEEauthorblockA{
        Confidential Draft \\
        Do NOT Distribute!!
      }
    }
  \fi 
}

\fancypagestyle{camerareadyfirstpage}{%
  \fancyhead{}
  
  \fancyhead[C]{
    \ifdefined\aeopen
    \parbox[][12mm][t]{13.5cm}{\hpcayear{} IEEE International Symposium on High-Performance Computer Architecture (HPCA)}    
    \else
      \ifdefined\aereviewed
      \parbox[][12mm][t]{13.5cm}{\hpcayear{} IEEE International Symposium on High-Performance Computer Architecture (HPCA)}
      \else
      \ifdefined\aereproduced
      \parbox[][12mm][t]{13.5cm}{\hpcayear{} IEEE International Symposium on High-Performance Computer Architecture (HPCA)}
      \else
      \parbox[][0mm][t]{13.5cm}{\hpcayear{} IEEE International Symposium on High-Performance Computer Architecture (HPCA)}
    \fi 
    \fi 
    \fi 
    \ifdefined\aeopen 
      \includegraphics[width=12mm,height=12mm]{ae-badges/open-research-objects.pdf}
    \fi 
    \ifdefined\aereviewed
      \includegraphics[width=12mm,height=12mm]{ae-badges/research-objects-reviewed.pdf}
    \fi 
    \ifdefined\aereproduced
      \includegraphics[width=12mm,height=12mm]{ae-badges/results-reproduced.pdf}
    \fi
  }
  \fancyfoot[C]{}
}
\begin{document}
\maketitle

\ifdefined\hpcacameraready 
  \thispagestyle{camerareadyfirstpage}
  \pagestyle{empty}
\else
  \thispagestyle{plain}
  \pagestyle{plain}
\fi

\newcommand{\hpcaheight}{0mm}
\ifdefined\eaopen
\renewcommand{\hpcaheight}{12mm}
\fi

\pagenumbering{arabic}

\begin{abstract}

Hardware prefetching plays a critical role in hiding the off-chip DRAM latency. The complexity of applications results in a wide variety of memory access patterns, prompting the development of numerous cache-prefetching algorithms. 
Consequently, commercial processors often employ a hybrid of these algorithms to enhance the overall prefetching performance. Nonetheless, \yao{since} these prefetchers share hardware resources, conflicts arising from competing prefetching requests can negate the benefits of hardware prefetching. Under such circumstances, several prefetcher selection algorithms have been proposed to mitigate conflicts between prefetchers. However, these prior solutions suffer from two limitations. First, the input demand request allocation is inaccurate. Second, the prefetcher selection criteria are coarse-grained. 


In this paper, we address both limitations by introducing an efficient \mm{and widely applicable} prefetcher selection algorithm---Alecto\footnote{The name \emph{Alecto} stands for the combination of \emph{selection} and \emph{allocation}.}, which tailors the demand requests for each prefetcher. Every demand request is first sent to Alecto to identify suitable prefetchers before being routed to prefetchers for training and prefetching. 
\mm{Our analysis shows that Alecto is adept at not only harmonizing prefetching accuracy, coverage, and timeliness but also significantly enhancing the utilization of the prefetcher table, which is vital for temporal prefetching.} 
Alecto outperforms the state-of-the-art RL-based prefetcher selection algorithm---Bandit by 2.76\% in single-core, and \mmreb{7.56\%} in eight-core.
\qijunre{For memory-intensive benchmarks, Alecto outperforms Bandit by 5.25\%.} Alecto consistently delivers state-of-the-art performance in scheduling various types of cache prefetchers. In addition to the performance improvement, Alecto can reduce the energy consumption associated with accessing the prefetchers’ table by \mm{48\%} \mmreb{(7\% energy reduction on the entire memory hierarchy)}, while only adding less than 1~KB of storage overhead.

\end{abstract}

\section{Introduction}
\label{sec:intro}

Hardware prefetching, a well-known technique to mitigate the ``memory wall'' \cite{wulf1995hitting}, has been exhaustively studied to improve processor performance. Specific data structures and data manipulation modes inherent in programs lead to diverse memory access patterns associated with the memory request addresses \cite{ayers2020classifying}. To efficiently predict these patterns, numerous types of cache prefetchers have been developed \cite{bakhshalipour2019evaluation, mittal2016survey, jouppi1990improving, he2022dsdp, hur2006memory, baer1991effective, dahlgren1995effectiveness, kim1997stride, somogyi2006spatial, kim2016path, michaud2016best, navarro2022berti, bakhshalipour2019bingo, bera2019dspatch, pugsley2014sandbox, shevgoor2015efficiently, ishii2009access, jiang2022merging, jain2013linearizing, nesbit2004data, wu2019efficient, wenisch2009practical, wu2019temporal, bakhshalipour2018domino, panda2023clip}. The most common among these are stream prefetchers \cite{jouppi1990improving, he2022dsdp, hur2006memory} and stride prefetchers \cite{baer1991effective, dahlgren1995effectiveness, kim1997stride}, which are leveraged to address stream and stride patterns respectively. To address more complex patterns, spatial prefetchers \cite{somogyi2006spatial, kim2016path, michaud2016best, navarro2022berti, bakhshalipour2019bingo, bera2019dspatch, pugsley2014sandbox, shevgoor2015efficiently, ishii2009access, jiang2022merging} and temporal prefetchers \cite{jain2013linearizing, nesbit2004data, wu2019efficient, wenisch2009practical, wu2019temporal, bakhshalipour2018domino} have been developed. 


No single prefetching algorithm can address all types of memory access patterns. In commercial processors, designers often hybrid several types of prefetchers, each specialized for specific memory access patterns. Ideally, these prefetchers should function independently, each concentrating on accurately addressing the specific memory access patterns they are responsible for \cite{kondguli2018division}. By doing so, the overall prefetching system can achieve maximal coverage of demand misses while simultaneously maintaining optimal accuracy. However, it is challenging to achieve this ideal separation among prefetchers. \yaoreb{First}, these prefetchers accept the same upstream dataflow, such as demand requests as training data. \mm{If a prefetcher receives demand requests \yaofix{beyond} its responsibility, it may (1) waste the precious space of prefetcher tables to store the associated metadata; (2) generate duplicate prefetching requests already produced by other prefetchers. Both of them affect the overall prefetching performance.} \yaoreb{Second}, prefetchers share internal or downstream hardware resources like prefetch table \cite{pakalapati2020bouquet}, prefetch queue, cache space, and DRAM bandwidth. This sharing can result in conflicts among the sub-prefetchers, where a prefetcher's occupation in such hardware resources slows down the prefetching progress of other prefetchers.


\begin{figure}[t]
\centering
\includegraphics[width=.99\linewidth]{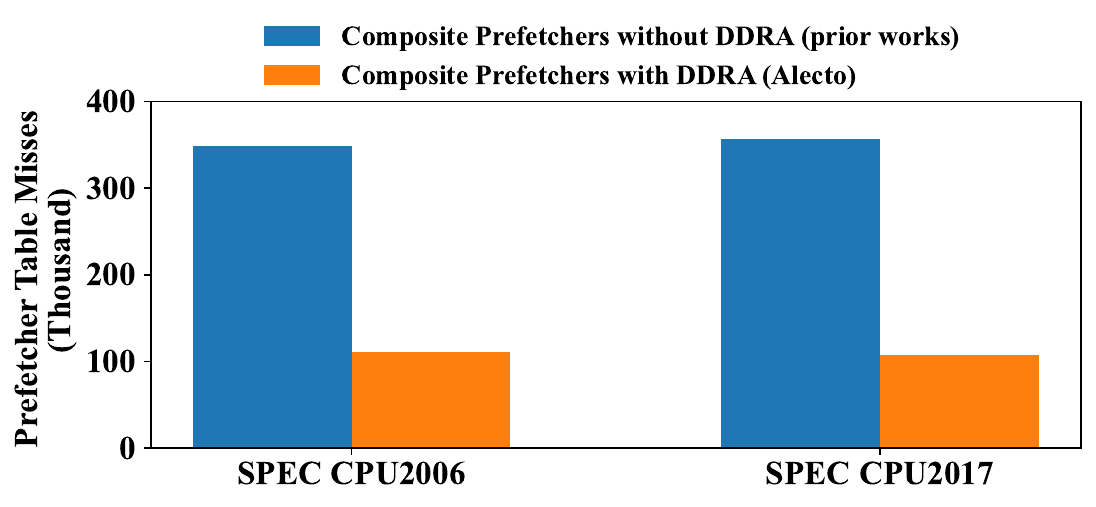}
\vspace{-.3in}
\caption{Comparison of prefetcher table misses in the \yaofix{same} \mm{composite prefetchers without dynamic demand request allocation (DDRA) and Alecto that utilizes DDRA.} With efficient demand request allocation, Alecto proves to significantly reduce conflicts that occur within the prefetchers’ table. }
\vspace{-.3in}
\label{fig:table_miss}
\end{figure}

These issues will not only \yao{impair} the benefits \yao{of hybrid} prefetchers but also affect system performance. To mitigate these side effects, several approaches for coordinating prefetchers have been proposed \cite{bera2021pythia, gerogiannis2023micro, alcorta2023lightweight, zhang2022resemble, pakalapati2020bouquet, kondguli2018division}. Their core idea is to identify and select suitable prefetchers for handling incoming demand requests. Suitable prefetchers should cover subsequent demand requests as much as possible while ensuring that the prefetching requests they issue are both accurate and timely. These strategies, however, fall short of expectations for two \yao{major limitations}. 

\def\thefootnote{$\dagger$}\footnotetext{Equal Contribution}\def\thefootnote{\arabic{footnote}}
\def\thefootnote{*}\footnotetext{Corresponding Author}\def\thefootnote{\arabic{footnote}}


\textbf{\yao{Limitation 1}: \yao{The input} \ming{demand request allocation} is \ming{inaccurate}.} Existing prefetcher selection algorithms mainly focus on controlling the prefetchers' output, such as the prefetching degree and prefetcher's on/off status. Though these regulations can mitigate some inaccurate prefetches and hardware conflicts, they do not prevent input demand requests \yao{training the tables of irrelevant prefetchers}. DOL~\cite{kondguli2018division} is the sole solution that selectively uses demand requests to train prefetchers. Nonetheless, it employs a static priority for all demand request allocations, without taking the difference of programs into account. 
\mm{Under existing solutions, prefetcher tables are often updated inaccurately due to receiving inappropriate demand requests. This leads to the replacement of original, potentially useful table entries, eventually causing an increase in the prefetcher table miss rate. 
Figure~\ref{fig:table_miss} presents a comparison of prefetcher table misses between prior works and our prefetcher selection algorithm\footnote{Our algorithm only allocates input demand requests to suitable prefetchers.} for scheduling the same composite prefetchers.} The results show that \yao{an appropriate input allocation} can significantly reduce the conflicts that occur within the prefetchers' table.

\begin{figure}[t]
\centering
\includegraphics[width=.99\linewidth]{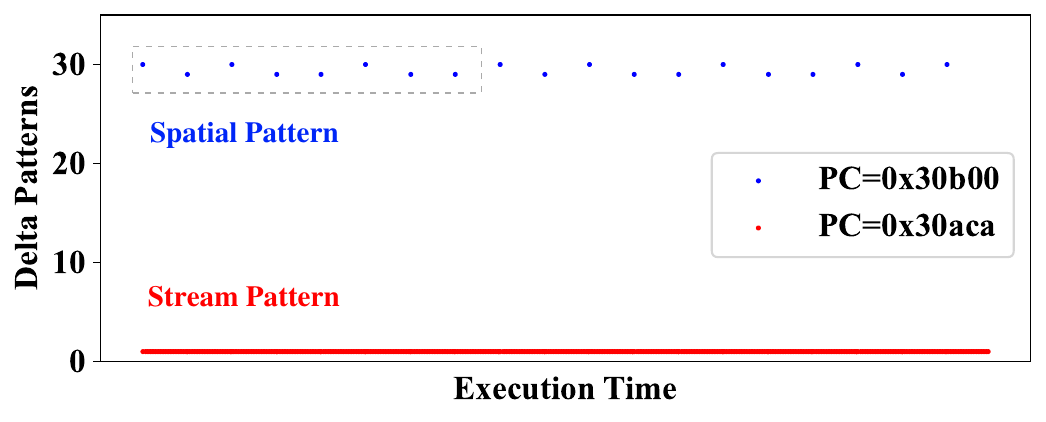}
\vspace{-.3in}
\caption{\mm{Memory access patterns of 459.GemsFDTD.} }
\vspace{-.2in}
\label{fig:data_point}
\end{figure}

\textbf{\yao{Limitation 2}: The \yao{prefetcher selection criteria} are coarse-grained.} Existing prefetcher selection algorithms apply identical \ming{selection rules} across varying demand requests. \mm{Figure~\ref{fig:data_point} shows an example of memory access patterns, taken from 459.GemsFDTD. \yaofix{It cannot be correctly handled by prior prefetcher selection algorithms}. In this example, two distinct memory access patterns are interleaved over time. The patterns from PC=0x30b00 should be handled by the spatial prefetcher (e.g., SPP \cite{kim2016path}) but could be erroneously \yaofix{handled} by the stride prefetcher. Prior works \cite{pakalapati2020bouquet, kondguli2018division} that use a uniform rule to select prefetchers for all incoming demand requests might prioritize the stride prefetcher over the spatial prefetcher for PC=0x30b00. Moreover, previous works \cite{gerogiannis2023micro} that establish the selection rules at runtime, without having PC-grained identification of suitable prefetchers, may prioritize the stream prefetcher for predicting PC=0x30aca, mistakenly treating memory accesses generated by PC=0x30b00 as noise.}

\mm{In this paper,} we propose a \yaofix{general design principle named dynamic demand request allocation (DDRA)}\,---\,allocating demand requests to suitable prefetchers to make each prefetcher get the best training and minimize conflicts within prefetcher tables. \yaofix{Based on the DDRA principle, we develop \texttt{Alecto}, an efficient prefetcher selection framework applicable for architecture with multiple prefetchers, overcoming the aforementioned limitations.} Alecto can dynamically allocate demand requests to suitable prefetchers. Upon identifying the suitable prefetchers of a given memory access, Alecto \yao{only} routes the demand requests to \yao{these suitable ones}, preventing the pollution of demand requests across \yao{irrelevant} prefetchers.



The dynamic demand request allocation in Alecto is supported by its ability to \textbf{pinpoint suitable prefetchers with fine-grained precision}. \ming{Alecto leverages the \mm{locally collected} performance \yao{record} of every prefetcher on each memory access instruction to determine the rules for subsequent prefetcher selections. The rules are tailored for each memory access instruction, specifying which prefetchers are eligible to receive the demand requests and their prefetching degree. }

In summary, \texttt{Alecto} integrates \emph{prefetcher selection} with \emph{dynamic demand request allocation}. For most of the cache prefetcher algorithms, the generation of prefetching requests is inherently linked to the training process of prefetchers. By controlling which demand request a prefetcher can access, the algorithm indirectly determines the outputs of prefetchers. The contributions of this paper are summarized below:




\begin{itemize}
    \item We thoroughly investigated existing prefetcher selection algorithms and pinpointed two major limitations. Our analysis encompasses a detailed review of all current solutions, and we illustrate why each fails to effectively address the critical issues inherent in prefetcher selection.
    \item \ming{We address \yao{limitations} of existing prefetcher selection algorithms in our proposed Alecto. Alecto \mm{is designed to enhance any multi-prefetcher architecture, introducing} three improvements over previous works: (1) dynamic demand request allocation; (2) fine-grained suitable prefetcher identification; (3) integration of prefetcher selection and demand request allocation.}
    \item \ming{Alecto \yao{delivers state-of-the-art performance, energy efficiency, and storage scalability} in a single, integrated framework. Extensive evaluations demonstrate Alecto outperforms Bandit by \mm{2.76\%} in single-core, \mmreb{7.56\%} in eight-core, and 5.25\% on memory-intensive benchmarks. Alecto offers broad adaptability in scheduling various types of cache prefetchers. Additionally, Alecto reduces energy consumption \yao{for prefetcher table access} by \mm{48\%} \mmreb{(7\% energy reduction on the entire memory hierarchy)}, with storage overhead $<$ 1~KB.}

    
\end{itemize}

\section{Background and Motivation}

\begin{figure*}[t]
\vspace{-.1in}
\hspace{-.4in}
\centering
\includegraphics[width=0.96\linewidth]{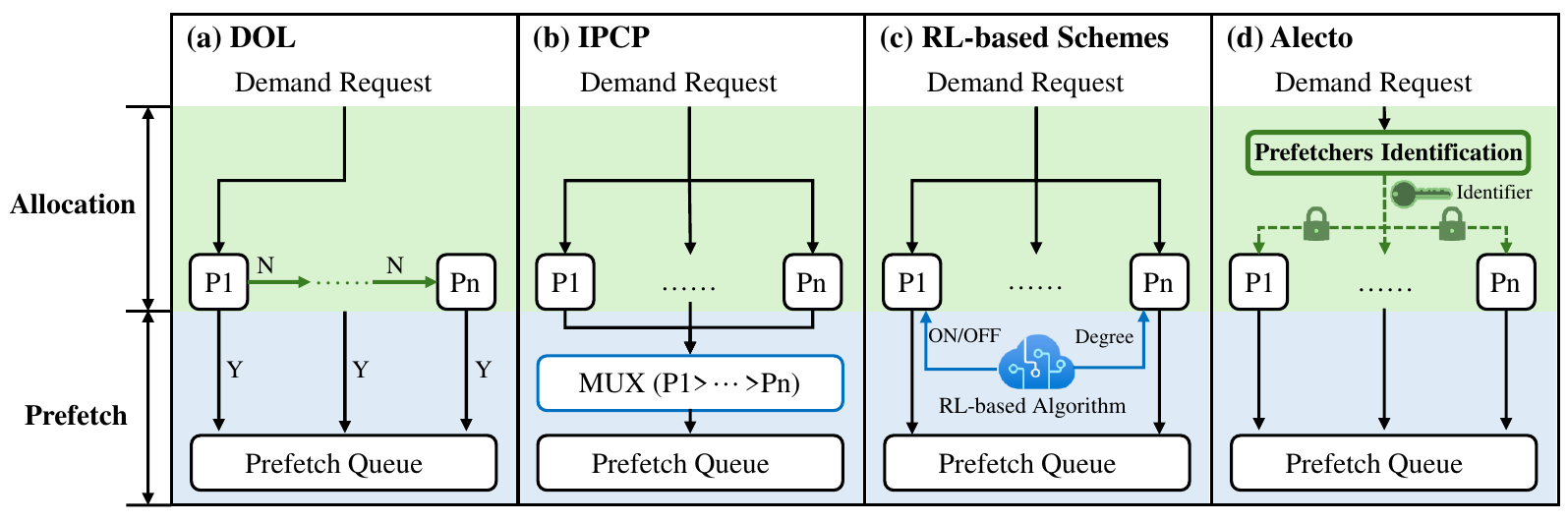}
\vspace{-.1in}
\caption{Comparison of prefetcher selection algorithms. (a) DOL selects prefetchers in the allocation stage. It sequentially passes the demand request through all prefetchers. (b) IPCP selects prefetchers in the prefetch stage. It statically prioritizes the prefetching requests from different prefetchers. (c) RL-based schemes select prefetchers in the prefetch stage. It controls the outputs of prefetchers and applies identical rules for all memory accesses. (d) Alecto selects prefetchers in the allocation stage. It identifies suitable prefetchers for each memory access, then dynamically allocates demand requests to identified prefetchers.}
\vspace{-.22in}
\label{fig:cmp}
\end{figure*}

In Section~\ref{subsec:solution}, we will analyze existing prefetcher selection solutions, illustrating why these approaches are inefficient in selecting suitable prefetchers when confronted with diverse memory access patterns. In Section~\ref{subsec:goal}, we will outline the objectives of our efficient and universally applicable prefetcher selection framework that addresses these shortcomings.

\subsection{Current Solutions}
\label{subsec:solution}

Figure~\ref{fig:cmp} compares different prefetcher selection algorithms, including the DOL \cite{kondguli2018division}, IPCP \cite{pakalapati2020bouquet}, and RL-based schemes \cite{bera2021pythia, gerogiannis2023micro, alcorta2023lightweight, zhang2022resemble}. DOL and IPCP utilize strategies established at design time, statically prioritizing prefetchers for handling every demand request, which is inefficient in selecting suitable prefetchers. While RL-based schemes hold promise for addressing this issue, they are challenged by the storage overhead for precise prefetcher suitability assessment. Additionally, all existing prefetcher selection solutions lack mechanisms for efficient demand request allocation. They either employ all demand requests for training all prefetchers like IPCP and RL-based algorithms, or statically pass demand requests through prefetchers such as DOL. Subsequent sections will illustrate these issues in detail.


\noindent \textbf{(1) DOL \cite{kondguli2018division}} employs a coordinator to determine which prefetcher each demand request is dispatched to. The coordinator, based on the prefetching coverage, sequentially passes demand requests through all prefetchers. For example, it operates under the assumption that the coverage of prefetcher P1 (T2 in the original paper) exceeds that of prefetcher P2 (P1 in the original paper). Therefore, as illustrated in Figure~\ref{fig:cmp}(a), demand requests are initially routed to P1. Only if P1 is unable to handle the demand request, is it then forwarded to P2, followed by P3 (C1 in the original paper).

DOL faces inefficiencies in selecting prefetchers primarily due to two reasons. First, statical priorities applied in allocating demand requests are infeasible in identifying the most suitable prefetchers for distinct demand requests. For instance, consider a scenario where a PC exhibits a consistent stride pattern of 1 (+1, +1, +1, +1). This pattern could potentially be identified as both a \texttt{StridePC}, serviced by prefetcher P1, and a \texttt{DensePC}, serviced by prefetcher P3. However, due to its predetermined priorities, DOL rigidly chooses P1 and omits P3. Despite P1 and P3 demonstrating equivalent accuracy and coverage, P3 offers better timeliness than P1. This is because P3 can prefetch an entire region at once, whereas P1 is limited to prefetching a single block. As a result, DOL should have selected P3 instead of P1 for higher timeliness yet it defaults to P1 because of its static selection approach. Second, DOL does not efficiently manage demand requests once the prefetchers for a specific pattern have been identified. DOL sequentially passes demand requests to P1, P2, and P3. One request might be best suited to P3, but its associated metadata (e.g., memory address and PC) could inadvertently leave traces in table structures of P1 and P2. This unintended recording can lead to the replacement of useful table entries in P1 and P2.

\noindent \textbf{(2) IPCP \cite{pakalapati2020bouquet}} hybrids three distinct types of prefetchers: P1 (GS, for stream patterns), P2 (CS, for stride patterns), and P3 (CPLX, for irregular patterns). As Figure~\ref{fig:cmp}(b) shows, these prefetchers accept all demand requests from the CPU core or higher-level caches as training data, evaluating their suitability for each request in parallel. When a single demand request could be serviced by more than one prefetcher, IPCP implements a static strategy to select the output of prefetchers based on a predetermined priority: P1 $>$ P2 $>$ P3.

Unlike DOL, IPCP primarily focuses on prioritizing the outputs of all prefetchers without specifically allocating demand requests to targeted prefetchers. This means every prefetcher observes the same dataflow, formed by the incoming demand requests. Consequently, each demand request contributes to updating the tables within IPCP. This \emph{non-selective} training can lead to more conflicts in prefetcher tables and reduce the duration that valuable table entries remain resident, diminishing the overall prefetching effectiveness. In addition, IPCP's inflexible priority system cannot identify suitable prefetchers like DOL. Consider an example where the PC displays a sequence of strides (+1, +1, +1, +4), which could be precisely predicted by P3. However, upon receiving the final (+1) stride, P2 incorrectly predicts this pattern, leading to an erroneous +1 request. The accurate requests, ideally produced by P3, are dropped due to the prioritization system.

\noindent \textbf{(3) RL-based Schemes \cite{gerogiannis2023micro, alcorta2023lightweight, zhang2022resemble}} leverage Reinforcement Learning (RL) to guide prefetcher decisions. One representative scheme is Bandit \cite{gerogiannis2023micro}, as depicted in Figure~\ref{fig:cmp}(c). Bandit employs an online learning strategy, which accepts the number of committed instructions as a reward, to control the degree of prefetchers. It turns off/on any prefetcher by crafting the prefetcher's degree to either zero or non-zero.

While Bandit represents an advancement over DOL and IPCP, it suffers from scalability issues that hinder it from identifying suitable prefetchers in a fine-grained manner and dynamically allocating demand requests. Its principal enhancement lies in the dynamic determination of prefetchers' outputs, with decisions driven by runtime metrics instead of static, predefined rules. This adaptability allows prefetchers to adjust to complex and varying memory access patterns more effectively. However, Bandit and other RL-based prefetcher selection schemes have two major defects. \qijunre{First, the storage overhead associated with scaling these algorithms---for instance, by increasing the number of features or actions is non-negligible, even though lightweight solutions like Bandit do not utilize ``state'' in their RL algorithm.} \mm{Bandit employs $\#actions^{\#prefetcher}$ arms (Section~\ref{subsec:storage}) for controlling each prefetcher's degree.} This could lead to an exponential increase in storage overhead. Second, existing RL-based schemes lack mechanisms for allocating demand requests. Integrating such functionality into RL-based frameworks is also challenging, as these algorithms cannot distinguish different demand requests.

\subsection{Our Goal}
\label{subsec:goal}


As shown in Figure~\ref{fig:cmp}(d), our goal is to develop a prefetcher selection framework that meets the following conditions:
\begin{itemize}
    \item \yaofix{The algorithm identifies the most suitable prefetchers for each memory access instruction. With this level of fine-grained precision, it tailors different prefetching strategies to various memory access instructions, optimizing the overall prefetching accuracy, coverage, and timeliness based on the unique characteristics of each request.} 
    \item The algorithm dynamically allocates demand requests to the identified prefetchers, ensuring that each prefetcher processes only the demand requests that fall within their designated fields. This targeted allocation not only provides each prefetcher with the specific training it needs but also helps to reduce conflicts within prefetcher tables. 
\end{itemize}

\section{Design Overview}
\label{sec:design}



In this section, we present Alecto, a scalable and efficient prefetcher selection strategy that utilizes runtime information for the fine-grained identification of suitable prefetchers and the dynamic allocation of demand requests. Alecto integrates the selection of prefetchers into the process of demand request allocation, achieving the goals stated in Section~\ref{subsec:goal}.

\subsection{Core Idea}

Since cache prefetchers depend on demand requests to train their tables, Alecto allocates each demand request to the identified suitable prefetchers, ensuring that every prefetcher receives the necessary training. This allocation process requires finely identifying the suitable prefetchers for each specific demand request. We have observed that demand requests originating from a single memory access instruction often display consistent patterns. Consequently, Alecto utilizes the associated PC information from incoming demand requests to pinpoint suitable prefetchers for these patterns. Specifically, Alecto leverages historical performance data of prefetchers to assess their suitability for specific memory instructions. In our implementation, the suitability assessment is based on the
prefetching accuracy. By focusing on accuracy, Alecto aims to make each prefetcher receive demand requests that fall within their designated field and reduce hardware resource wastage.

\subsection{Structure Overview}
\label{subsec:overview}

\begin{figure}[t]
\centering
\includegraphics[width=\linewidth]{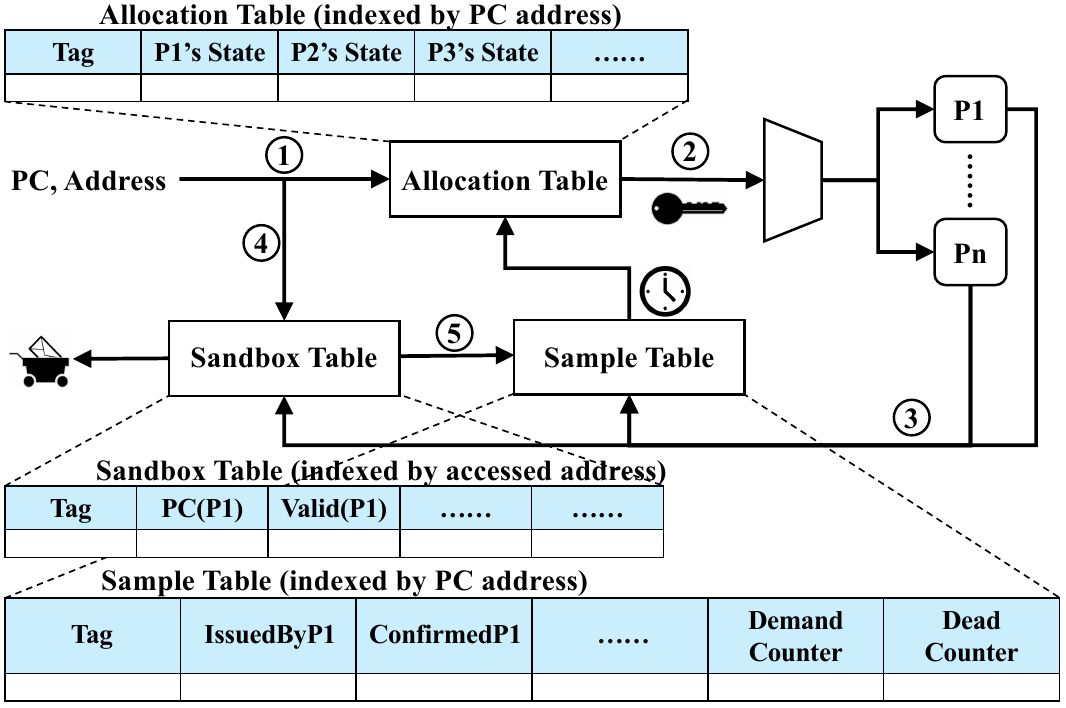}
\vspace{-.15in}
\caption{\mmreb{The overall framework of Alecto. It consists of an Allocation Table, which enables fine-grained prefetcher identification and dynamic request allocation. It also includes a Sample Table and Sandbox Table for information collection. Additionally, the Sandbox Table functions as a prefetch filter. }}
\vspace{-.25in}
\label{fig:framework}
\end{figure}

Figure~\ref{fig:framework} outlines the primary structures of Alecto. It primarily encompasses three hardware components:

\begin{itemize}
    \item \textit{Allocation Table}: This table is the foundational component of Alecto. It facilitates the fine-grained identification of suitable prefetchers (Section~\ref{subsec:identification}) by documenting the state of each prefetcher for every memory access instruction. These states, informed by each prefetcher's historical performance, reflect the efficacy of prefetchers in handling demand requests generated by the memory access instruction. Based on these states, this table can dynamically allocate demand requests to suitable prefetchers (Section~\ref{subsec:alloc}).
    \item \textit{Sample Table}: This table collects runtime information and periodically forwards the accumulated data to the Allocation Table, aiding in the process of updating the states of prefetchers (Section~\ref{subsec:gather}).
    \item \textit{Sandbox Table}: This table serves dual functions: (1) recording recently issued prefetching requests and collecting useful prefetches for the Sample Table (Section~\ref{subsec:gather}), and (2) acting as a prefetch filter to eliminate duplicate prefetching requests (Section~\ref{subsec:filter}).
\end{itemize}

\subsection{Process Overview}
At the beginning of our framework, the demand request, including the PC and memory address, is simultaneously sent to the Allocation Table (step \ding{172}) and the Sandbox Table (step \ding{175}). According to the PC and documented states, the Allocation Table generates an \textit{identifier} to instruct the allocation of demand requests (step \ding{173}). Only prefetchers matched with the identifier can accept this request and accordingly generate prefetching requests. Subsequently, feedback from these selected prefetchers---in the form of issued prefetching requests---is utilized to update the Sandbox Table and Sample Table (step \ding{174}). Combined with the step \ding{175}, the Sandbox Table can make certain whether a previous prefetching request is useful. This information is sent to the Sample Table (step \ding{176}), enabling it to derive a prefetcher's accuracy for each memory access instruction. Finally, the Sandbox Table is extended as a prefetch filter, intercepting redundant prefetch requests and routing the unfiltered requests to next-level caches (step \ding{177}).

\section{Design Implementation}

\subsection{Fine-grained Prefetchers Identification}
\label{subsec:identification}


\begin{figure}[t]
\vspace{-.1in}
\centering
\includegraphics[width=\linewidth]{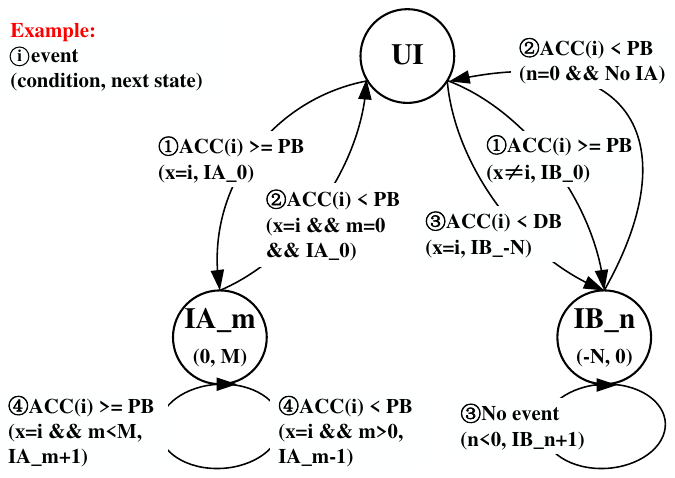}
\vspace{-.25in}
\caption{\mmreb{The state machine of Allocation Table. For every memory access instruction, each prefetcher has three states: Un-Identified (UI) indicates the suitability of this prefetcher is unidentified; Identified and Aggressive (IA) means the prefetcher is efficient and its prefetching degree should be promoted; Identified and Blocked (IB) applies when a prefetcher is deemed unsuitable for processing the memory access instructions.}}
\vspace{-.2in}
\label{fig:state_machine}
\end{figure}

Alecto utilizes the Allocation Table to identify suitable and unsuitable prefetchers for each memory access instruction, guiding the allocation of demand requests to the identified prefetchers. Figure~\ref{fig:framework} depicts the structure of the Allocation Table, which is indexed by the address of memory access instructions (PCs) and stores the state of each prefetcher. The state information is critical for Alecto's decision-making process regarding the allocation of demand requests and the regulation of prefetcher aggressiveness. To facilitate this, Alecto designates three states for every prefetcher: the \textbf{Un-Identified (UI)} state, \textbf{Identified and Aggressive (IA)} state, \textbf{Identified and Blocked (IB)} state. Moreover, Alecto introduces two thresholds to guide state transitions: the \textbf{Proficiency Boundary (PB)} and \textbf{Deficiency Boundary (DB)}. Metrics gathered from the Sample Table are juxtaposed against these thresholds to ascertain the appropriate state. Alecto adheres to \textbf{two principles} when applying these states and boundaries to the identification of suitable prefetchers: (1) High-performing prefetchers are prioritized by allocating demand requests to them and withholding these requests for other prefetchers. Concurrently, these high-performing prefetchers are advanced to a higher level of aggressiveness, optimizing their performance further. (2) Underperforming prefetchers are temporarily restricted from receiving demand requests to mitigate inefficiency. Recognizing that PC's memory access pattern changes over time, these prefetchers are only suspended for a limited duration rather than being permanently deactivated, allowing for future reevaluation and adjustment.

The ensuing discussion in this section will delve into the significance of each state and elucidate how Alecto employs these states in the selection process of prefetchers.

Firstly, we explain the meaning of each state:

\begin{itemize}
    \item \textbf{UI} State: This state signifies a prefetcher's suitability is undetermined. Under this state, Alecto employs a prudent approach by allocating demand requests to the prefetcher while limiting its prefetching degree to a conservative level, such as two.
    
    \item \textbf{IA} State: This state indicates a prefetcher is efficient. Alecto allocates demand requests to this prefetcher and elevates its prefetching degree. It encompasses M+1 substates, numbered from 0 to M, with higher values indicating a larger prefetching degree assigned for the prefetcher. 
    
    \item \textbf{IB} State: This state applies when a prefetcher is deemed unsuitable for processing the memory access instruction. Alecto closes the flow of demand requests to the prefetcher. It has N+1 sub-states, ranging from -N to 0. The smaller the value, the longer time the prefetcher will be blocked.
    
\end{itemize}

Upon a new, unrecorded memory access instruction enters into the Allocation Table, the initial state for all prefetchers is set to \textbf{UI} until the Sandbox Table and Sample Table accumulate sufficient data to determine the suitability of prefetchers. In our implementation, we use these two tables to calculate the prefetching accuracy. To quantify what constitutes `sufficient data,' Alecto defines a timing epoch based on the volume of demand accesses it encounters. According to our observations, an epoch marked by 100 demand accesses is adequate. The possible state transitions within the Allocation Table are illustrated in Figure~\ref{fig:state_machine}:

\begin{itemize}
    \item \textbf{\ding{172} (UI$\,\to\,$IA\_0, UI$\,\to\,$IB\_0)} occurs when the accuracy of one or more prefetchers surpasses the PB. Prefetchers exceeding this accuracy threshold are elevated to the \textbf{IA\_0} state. Conversely, the remaining prefetchers that do not meet this criterion are transitioned to \textbf{IB\_0}. One exception occurs when two or more prefetchers are considered for promotion, and one of them is a temporal prefetcher. In such cases, Alecto promotes the non-temporal prefetchers and downgrades the temporal prefetcher. This approach aims to optimize the metadata storage of the temporal prefetcher. (Section~\ref{subsec:temporal})
    \item \textbf{\ding{173} (IA\_0$\,\to\,$UI)} arises when prefetchers, previously in the \textbf{IA\_0} state, experience a drop in accuracy below the PB. In such cases, Alecto downgrades the prefetchers back to the \textbf{UI} state, reflecting uncertainty in their current suitability. If this action results in the absence of any prefetcher in the \textbf{IA} state, Alecto further initiates the transition of prefetchers from \textbf{IB\_0} to \textbf{UI} \textbf{(IB\_0$\,\to\,$UI)}. \mm{This process facilitates a reassessment of the prefetchers that have cooled down from lower IB\_n states to IB\_0.}
    \item \textbf{\ding{174} (UI$\,\to\,$IB\_-N)} is triggered when a prefetcher's accuracy falls below the DB. This situation indicates that the prefetcher has generated an excessive number of inaccurate prefetches, wasting valuable hardware resources. In response, Alecto demotes the prefetcher's state to \textbf{IB\_-N}, signifying that this prefetcher should be temporarily blocked for N epochs due to its inefficiency. During these N epochs, the prefetcher undergoes a gradual transition (\textbf{IB\_n$\,\to\,$IB\_n+1}), incrementally moving towards reevaluation. At the end of this process, if no prefetcher remains in the \textbf{IA} state, the blocked prefetcher may be reconsidered for promotion back to \textbf{UI} state, allowing it another opportunity to evaluate its suitability.
    \item \textbf{\ding{175} (IA\_m$\,\to\,$IA\_m+1, IA\_m$\,\to\,$IA\_m-1)} takes place when the accuracy of prefetchers in \textbf{IA\_m} state surpasses the PB or dips below the DB. For the former case, we infer that these prefetchers possess the potential to cover more cache misses and achieve better timeliness. Therefore, Alecto elevates their state to \textbf{IA\_m+1}, granting them an increased prefetching degree. The opposite state transition adjusts the operations accordingly.
\end{itemize}


\subsection{Dynamical Demand Request Allocation}
\label{subsec:alloc}


When demand requests enter into the Allocation Table, Alecto conducts a lookup using the PC to identify prefetchers currently in the \textbf{UI} and \textbf{IA\_m} states. If a prefetcher belongs to the \textbf{UI} state, Alecto generates an identifier that includes the prefetcher's sequence number and a conservative prefetching degree $c$. Conversely, when a prefetcher is positioned in the \textbf{IA\_m} state, Alecto pairs its sequence number with a more aggressive prefetching degree, calculated as $c+m+1$. To efficiently manage cache space, Alecto prefetches $c$ lines directly into the cache where the prefetchers reside. For the additional $m+1$ lines, Alecto prefetches them to the next-level cache. The identifiers are routed along with demand requests to a multiplexer, which then dynamically allocates demand requests to designated prefetchers based on these identifiers. Our demand request allocation methodology ensures that prefetchers in different states are appropriately tasked based on their demonstrated performance and potential for addressing demand requests.

\subsection{Runtime Metrics Gathering}
\label{subsec:gather}

\mm{Alecto utilizes two structures: the Sample Table, which is indexed by the PC address, and the Sandbox Table, indexed by the memory access address, to collect runtime metrics that are periodically forwarded to the Allocation Table.} The data within these tables are updated in receiving demand requests (step \ding{172} in Figure~\ref{fig:framework}) and issuing prefetching requests (step \ding{176} in Figure~\ref{fig:framework}). 

The Sample Table tracks the total number of prefetching requests issued by each prefetcher \mmreb{(``IssuedByP1'' in Figure~\ref{fig:state_machine})}, as well as the number of requests hit by subsequent demand requests \mmreb{(``ConfirmedP1'' in Figure~\ref{fig:state_machine})}. The identification of such hit information is facilitated by the Sandbox Table, which records the addresses of recently issued prefetching requests and the PC information. The PC indicates which memory access instruction triggers those prefetching requests. When a demand request aligns with an entry's tag in the Sandbox Table, Alecto then verifies if the PC of the demand request matches the recorded PC for any prefetcher. If a match is found, Alecto increments the confirmed counter in the Sample Table by one. To efficiently manage the storage of PCs within the Sandbox Table, Alecto utilizes common hash functions found in Branch Prediction Unit (BPU) designs \cite{hennessy2011computer, seznec2014tage, seznec2011new, seznec201164}. This approach involves dividing the PC address into n segments and applying an XOR operation across these segments to generate a final, compacted hash value for storage in the Sandbox Table. By setting n to correspond with the logarithm of the table's entry count, Alecto significantly decreases the storage overhead.


In addition to gathering data for calculating prefetching accuracy, the Sample Table plays a crucial role in tracking the number of demand requests encountered by Alecto (indicated as the ``Demand Counter" in Figure~\ref{fig:framework}) and the instances when Alecto does not issue any prefetch requests (``Dead Counter" in Figure~\ref{fig:framework}). The Demand Counter is instrumental in defining the timing epoch to update the Allocation Table. Assigning a separate Demand Counter (e.g., 100) for each PC ensures that sufficient information is collected before any adjustments are made to prefetcher behavior. \mmreb{Alecto resets the Demand Counter when it reaches its threshold, triggering the state transition of the associated PC.} Furthermore, the Sample Table employs the Dead Counter to prevent deadlock scenarios\mmreb{, which occur when a PC remains in the IA\_m state but fails to generate prefetch requests for a long period. These deadlock scenarios can arise when the memory access pattern associated with a PC transitions to a different type, rendering the previously identified pattern for that PC invalid. This counter operates as a saturation counter and is not reset along with the Demand Counter.} It increments each time Alecto fails to generate a prefetch request during a prediction and decreases in other situations. Reaching a specific threshold with the Dead Counter prompts Alecto to reset the states of prefetchers in the Allocation Table back to \textbf{UI}, initiating a fresh search for suitable prefetchers for the current memory access instruction. Experimental findings indicate a low probability of deadlock, leading to the threshold for the Dead Counter being set higher than that for the Demand Counter, for example, at 150.

\subsection{Duplicate Prefetch Requests Filtering}
\label{subsec:filter}

Alecto extends the functionality of the Sandbox Table to a prefetch filter, which is leveraged to filter out duplicate prefetching requests. As detailed in Section~\ref{subsec:gather}, this table logs the recently issued prefetch requests from each prefetcher, facilitating its role as an efficient prefetch filter. During step \ding{177}, Alecto queries the Sandbox Table with the address of prefetching requests. If a request matches an existing tag in the table (a tag hit), Alecto discards the request to prevent redundant prefetching. If there's no tag hit, the request is added to the prefetch queue, enabling the prefetching of data blocks into caches. This mechanism operates without imposing extra storage demands. Further, as discussed in Section~\ref{subsec:why}, the Sandbox Table is critical in simultaneously maximizing the prefetching accuracy and coverage.

\subsection{Why Alecto Achieves Our Goal?}
\label{subsec:why}

We demonstrate that the prefetchers selection strategies applied in Alecto achieve our goal mentioned in Section~\ref{subsec:goal}. Addressing the first requirement, Alecto assigns each memory access instruction to specific states within the Allocation Table. These states reflect the suitability of each prefetcher for handling that particular instruction. Based on the states, Alecto tailors prefetching strategies for maximizing the prefetching accuracy, coverage, and timeliness. Specifically, Alecto advances prefetchers exhibiting sufficient accuracy ($>$PB) to the \textbf{IA\_m} state, rather than only those with the absolute highest accuracy. This approach is designed to balance the overall accuracy and coverage. If we exclusively promote the prefetcher with the highest accuracy, there is a risk it may offer limited coverage. Alecto does not worry about prefetchers in \textbf{IA\_m} state generating duplicate requests. This is because, in the final phase of Alecto, the Sandbox Table acts as a filter to eliminate these redundant prefetching requests. Besides focusing on prefetching accuracy and coverage, Alecto incrementally adjusts the aggressiveness of prefetchers in the \textbf{IA\_m} state to a suitable level, i.e., either when accuracy falls below the PB or when the maximum degree of aggressiveness is reached. This approach not only maintains the accuracy and coverage but also enhances the timeliness of prefetchers.

For the second requirement, Alecto inherently integrates the process of prefetcher selection within the demand request allocation. This ensures that as demand requests are allocated, the suitable prefetchers are simultaneously selected. Specifically, prefetchers identified as unsuitable are maintained in the \textbf{IB\_n} state. As detailed in Section~\ref{subsec:alloc}, Alecto does not create identifiers for prefetchers in the \textbf{IB\_n} state, ensuring that demand requests are not allocated to these prefetchers. This mechanism intrinsically prevents unsuitable prefetchers from receiving data that could lead to unnecessary training or inaccurate prefetching.

\subsection{Alecto on Temporal Prefetching}
\label{subsec:temporal}

\begin{figure}[!t]
\centering
\includegraphics[width=\linewidth]{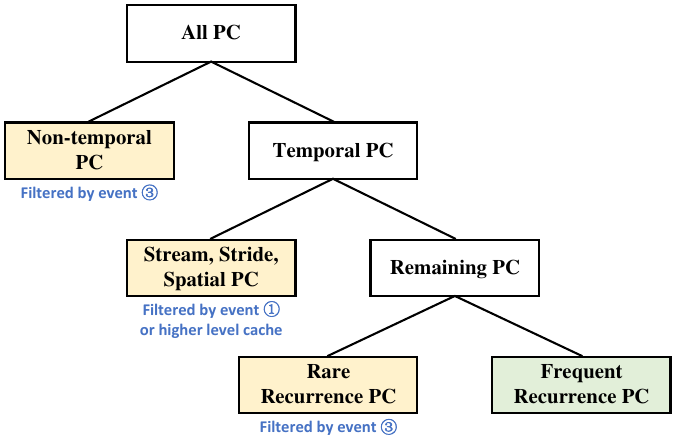}
\vspace{-.3in}
\caption{\mm{The classification of memory access patterns for efficient metadata storage in temporal prefetching. Alecto can filter demand requests that are (1) non-temporal; (2) simultaneously handled by non-temporal prefetchers; and (3) rare recurrence. }}
\vspace{-.25in}
\label{fig:tp}
\end{figure}

\mm{The dynamic demand request allocation technique adopted by Alecto can significantly enhance the utilization of temporal prefetcher's metadata storage. We find that only a small portion of demand requests should be used to train the table of temporal prefetchers. To illustrate this idea, we have divided the memory access patterns into several parts, as shown in Figure~\ref{fig:tp}. Firstly, non-temporal patterns, formed by a sequence of non-recurrence of memory accesses, should not train the temporal prefetcher. For temporal patterns, temporal prefetchers can logically process all such patterns. However, we can further improve the metadata storage utilization by filtering demand requests when (1) the recurring frequency of these demand requests is too low so that the prefetcher cannot preserve related metadata until the next recurrence; (2) these requests can be simultaneously handled by non-temporal prefetchers such as stream, stride, and spatial prefetchers with lower storage overhead.}

To the best of our knowledge, previous prefetcher selection algorithms \cite{bera2021pythia, gerogiannis2023micro, alcorta2023lightweight, zhang2022resemble, pakalapati2020bouquet, kondguli2018division} do not provide a mechanism for allocating demand requests to temporal prefetching. While the state-of-the-art temporal prefetcher---Triangel \cite{ainsworth2024triangel} can filter non-temporal PCs and rare recurrence PCs, it falls short in (1) lacking mechanism filtering PC handled by non-temporal prefetcher, (2) suffering from high storage overhead ($>$ 17KB), and (3) being tightly coupled with temporal prefetcher and failing to schedule other prefetchers.

Without any modification, Alecto serves as a lightweight ($<$ 1KB) but efficient solution that can filter all excessive demand requests for temporal prefetchers. Assume the system integrates all types of prefetchers for maximizing the prefetching chances, there are two scenarios:

\begin{itemize}
    \item \mm{\textbf{The temporal prefetcher and other non-temporal prefetchers are located at the same cache level}: Alecto can filter out the demand requests beyond the temporal prefetcher's capacity by event \ding{174} (Figure~\ref{fig:state_machine}) and those within non-temporal prefetchers' capabilities by event \ding{172}.}
    \item \mm{\textbf{They are located at different cache levels}: Non-temporal prefetchers cause demand requests belonging to the stream, stride, and spatial patterns hitting the higher level cache. Consequently, the temporal prefetcher only needs to filter demand requests outside its capacity, which can be done by event \ding{174}.}
\end{itemize}

\section{Experimental Methodology}
\label{sec:method}

\subsection{System Configuration}

\begin{table}[t]
\centering
\vspace{-5mm}
\caption{System Configuration.}
\vspace{-2mm}
\label{tab:config}
\begin{tabular}{|l|l|}
\hline
\textbf{Module} & \textbf{Configuration} \\
\hline
\hline 
Core & 1-8 cores, 256-entry ROB\\
&6-width fetch, 6-width decode \\
&8-width issue, 4-width commit \\
&256-entry IQ, 72/56-entry LQ/SQ \\
\hline
\mm{TLBs} & \mm{64-entry L1 iTLB/dTLB, 8-way}  \\
& \mm{2048-entry shared L2 TLB, 16-way} \\
\hline
Private L1 I/D cache & 32~KB each, 8-way, 64B line, 16 MSHRs  \\
&LRU, 4 cycles round-trip latency \\
\hline
Private L2 cache & 256~KB, 8-way, 64B line, 32 MSHRs \\
&LRU, mostly\_inclusive \\
&15 cycles round-trip latency \\
\hline
Shared L3 cache & 2~MB per core, 16-way, 64B line \\
&64 MSHRs per LLC Bank \\
&CHAR \cite{chaudhuri2012introducing}, mostly\_exclusive \\
&35 cycles round-trip latency \\
\hline
Main Memory 
& SC: Single channel, 1 rank/channel \\
& MC: $\frac{\#Core}{2}$ channels, 2 ranks/channel \\
& 8 banks/rank, 2400 MTPS \\
\hline
\end{tabular}
\vspace{1mm}
\vspace{-4mm}
\end{table}

For an in-depth examination of different prefetcher selection algorithms' performance, we utilize the execution-driven simulator gem5 \cite{binkert2011gem5}. Our simulation environment adopts parameters almost consistent with those utilized in the Bandit \cite{gerogiannis2023micro} studies, resembling the Intel Skylake processor \cite{skylake}. The primary system configurations are outlined in Table~\ref{tab:config}.

\subsection{Prefetchers Simulated}
\label{subsec:psim}

\begin{table}[b]
\centering
\vspace{-3mm}
\caption{\mmreb{Prefetchers Being Selected.}}
\vspace{-2mm}
\label{tab:sel}
\begin{tabular}{|l|l|}
\hline
\mmreb{\textbf{Component}} & \mmreb{\textbf{Configuration}} \\
\hline
\hline 
\mmreb{Stream prefetcher} & \mmreb{64-entry IP table} \\
\mmreb{(GS in \cite{pakalapati2020bouquet})} & \mmreb{8-entry Region Stream Table (RST)} \\
\hline
\mmreb{Stride prefetcher} & \mmreb{64-entry IP table}  \\
\mmreb{(CS in \cite{pakalapati2020bouquet})} & \\
\hline
\mmreb{Spatial prefetcher} & \mmreb{16-entry Accumulation Table}  \\ 
\mmreb{(PMP \cite{jiang2022merging})} & \mmreb{64-entry Pattern History Table (PHT)}  \\
\hline
\end{tabular}
\end{table}



We compare Alecto against \yaoreb{multiple existing prefetcher \emph{selection algorithms} from} IPCP \cite{pakalapati2020bouquet}, DOL \cite{kondguli2018division}, and Bandit \cite{gerogiannis2023micro}. \yaoreb{We apply all selection algorithms to schedule \emph{exactly the same} set of prefetchers like} Arm Neoverse V2 architecture \cite{arm-neov2}, containing a stream prefetcher (GS in \cite{pakalapati2020bouquet}), a stride prefetcher (CS in \cite{pakalapati2020bouquet}), and a spatial prefetcher (PMP \cite{jiang2022merging}).


For Alecto, we define N for the IB\_n state as 8, M for the IA\_m state as 5, \mm{c for the conservative prefetching degree as 3}, set the Proficiency Boundary (PB) at 0.75, and the Deficiency Boundary (DB) at 0.05. \mm{For IPCP, demand requests are distributed to all prefetchers. The output from these prefetchers is selected based on the static priority order: stream prefetcher $>$ stride prefetcher $>$ spatial prefetcher.} For DOL, demand requests are sequentially passed to stream, stride, and spatial prefetchers; the process halts when one successfully processes the data, preventing further distribution. \mm{For Bandit, we customize it to regulate the prefetching degree of prefetchers}. Bandit typically considers around 10 candidate arms. Given our three-prefetcher setup, we designate two states for the prefetching degree of each prefetcher---either 0 or X, where X is a configurable parameter. This configuration leads to $2^3=8$ candidate arms. In our evaluations, X is set to 3 (Bandit3) or 6 (Bandit6). 

For a fair comparison, all simulated prefetchers are implemented within the L1 data cache \mm{and trained using virtual address.} Considering Alecto naturally has a prefetch filter, we additionally add a prefetch filter for other configurations to better reflect real-world conditions.

\subsection{Evaluating Alecto on Temporal Prefetching}
\label{subsec:tp_config}

\begin{figure}[t]
\vspace{-.15in}
\centering
\includegraphics[width=\linewidth]{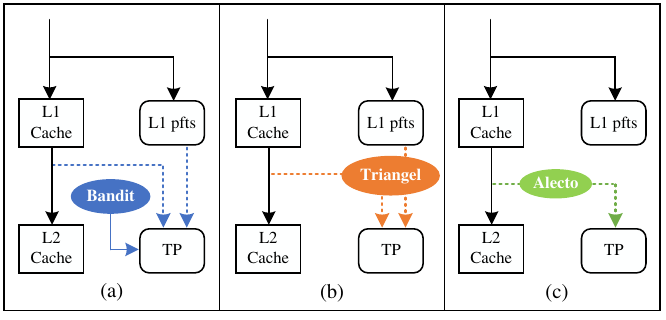}
\vspace{-.25in}
\caption{\mm{The comparison of configurations between (a) Bandit; (b) Triangel; and (c) Alecto on temporal prefetching (TP).}}
\vspace{-.25in}
\label{fig:tp_exp}
\end{figure}

We evaluate the effectiveness of Alecto on temporal prefetching by comparing it against Bandit and Triangel \cite{ainsworth2024triangel}. We implement the temporal prefetcher using an on-chip scheme like Triangel \cite{ainsworth2024triangel}. Figure~\ref{fig:tp_exp} displays the three configurations evaluated. Referring to previous works \cite{wu2021practical, wu2019temporal, wu2019efficient}, the Bandit and Triangel configuration has the temporal prefetcher training on every L2 access stream, which includes both prefetching requests from L1 prefetchers (IPCP) and demand requests from L1 cache. The difference is that Bandit only controls the prefetching degree, while Triangel can further manage the temporal prefetcher's training requests. Different from Bandit and Triangel, our solution only receives demand requests from the L1 cache and uses Alecto to manage them (Section~\ref{subsec:temporal}). \mmreb{In these three configurations, we adopt the same metadata format as in Triangle and set the maximum prefetching degree of the temporal prefetcher to 1. For Triangel, we use the code \cite{triangel} open-sourced by its original paper and preserve its complete implementation. But for Bandit and Alecto, we disable Triangel's metadata table management strategies to accurately evaluate the impact of demand request allocation on metadata table size. Both Alecto and Bandit are configured with a 1MB L3 cache and a 1MB metadata table.}

\subsection{Workloads}

\begin{figure*}[t!]
\centering
\includegraphics[width=.99\linewidth]{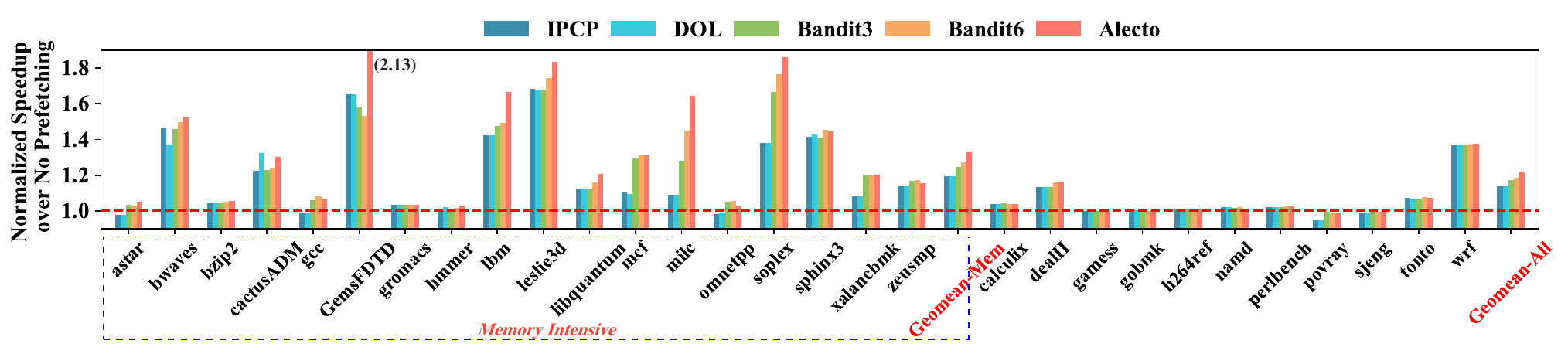}
\vspace{-.13in}
\caption{\mm{IPC speedup compared to no prefetching on SPEC06 benchmarks.} \mmreb{Memory-intensive workloads are highlighted within dotted blue line, and their geomean speedup is calculated separately. All prefetcher selections algorithms schedule the same composite prefetcher: GS \cite{pakalapati2020bouquet} + CS \cite{pakalapati2020bouquet} + PMP \cite{jiang2022merging}.}}
\vspace{-.15in}
\label{fig:spec06}
\end{figure*}

\begin{figure*}[t!]
\vspace{.05in}
\centering
\includegraphics[width=.99\linewidth]{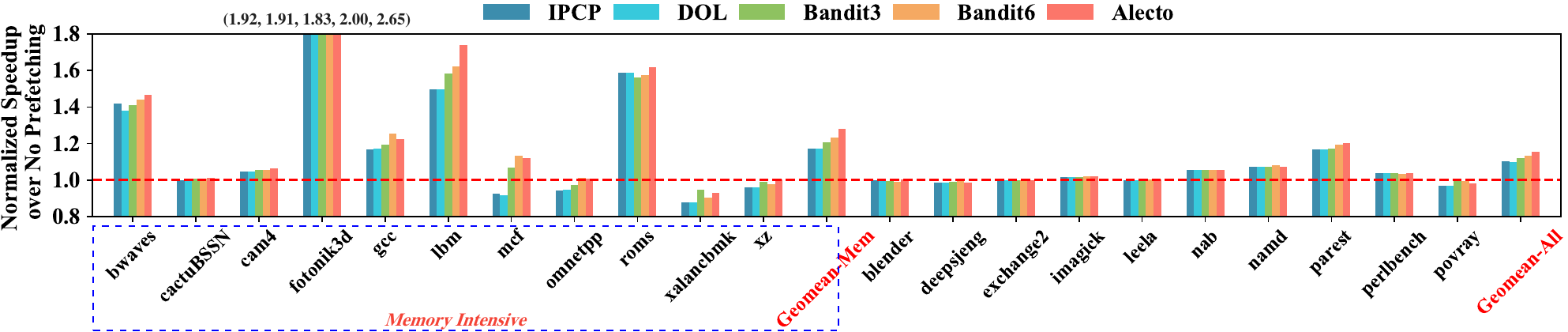}
\vspace{-.1in}
\caption{\mm{IPC speedup compared to no prefetching on SPEC17 benchmarks.} \mmreb{Memory-intensive workloads are highlighted within dotted blue line, and their geomean speedup is calculated separately. All prefetcher selections schemes algorithms the same composite prefetcher: GS \cite{pakalapati2020bouquet} + CS \cite{pakalapati2020bouquet} + PMP \cite{jiang2022merging}.}}
\vspace{-.25in}
\label{fig:spec17}
\end{figure*}

We evaluate Alecto on a broad spectrum of workloads, including the SPEC CPU2006 \cite{SPEC2006}, SPEC CPU2017 \cite{SPEC2017}, PARSEC 3.0 \cite{PARSEC}, and Ligra \cite{shun2013ligra}. For single-core simulation, we use the SPEC06 and SPEC17 benchmarks and adopt simpoint techniques \cite{sherwood2002automatically} for generating checkpoints across all SPEC benchmarks. In multi-core simulation, we use PARSEC and Ligra. For PARSEC, we concentrate on each benchmark's Region of Interest (ROI), which includes the parallel code in the program \cite{southern2016analysis, gebhart2009running}. We take checkpoints at the beginning of the ROI to speed up our evaluations. According to our observations, the simpoint and ROI techniques allow us to capture typical patterns of each benchmark.

In our single-core experiments, each checkpoint is pinned to an individual core, with a warm-up phase involving \mm{100M} instructions followed by a simulation of the subsequent \mm{100M} instructions. The reported performance metrics for each benchmark are calculated by aggregating the results from all its checkpoints with weighted averages. \mmreb{In our multi-core experiments, we evaluate homogeneous mixes and heterogeneous mixes of SPEC benchmarks. For the homogeneous mixes, we pin the same SPEC workload to every core. For the heterogeneous mixes, we randomly choose workloads from SPEC and pin them to different cores.} We also evaluate PARSEC and Ligra. For SPEC and PARSEC, each checkpoint is warmed up with \mmreb{250M} instructions, followed by simulation with an additional \mmreb{250M} instructions. For Ligra, we use the input file \texttt{rMatGraph\_WJ\_5\_100} across all tests and execute the entire benchmark.

\section{Evaluation}
\label{sec:eva}

\subsection{Single-core Evaluation}
\label{subsec:single}

\begin{figure}[t]
\centering
\includegraphics[width=\linewidth]{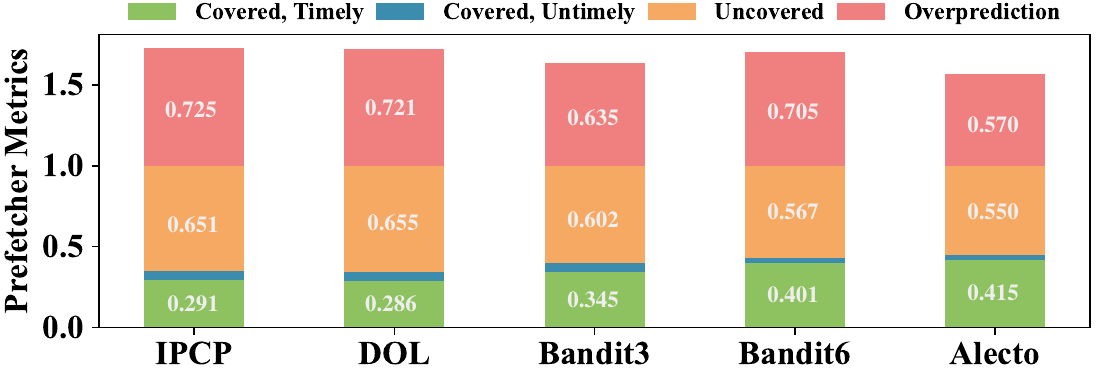}
\vspace{-.25in}
\caption{\mm{The key performance metrics of prefetchers.}}
\vspace{-.25in}
\label{fig:metric}
\end{figure}

Figure~\ref{fig:spec06} and Figure~\ref{fig:spec17} report the IPC speedup of all six evaluated prefetchers over a non-prefetching baseline, across the SPEC CPU2006 and SPEC CPU2017 benchmarks. For SPEC06 benchmarks, Alecto outperforms IPCP by \mm{8.14\%}, DOL by \mm{8.04\%}, Bandit3 by \mm{4.77\%}, Bandit6 by \mm{3.20\%}\footnote{\yaoreb{In all our evaluations, IPCP, DOL, Bandit only refer to prefetcher \emph{selection} algorithms. They schedule exactly the same composite prefetcher (e.g., GS+CS+PMP) as Alecto, rather than the prefetchers in their original papers.}}. For SPEC17 benchmarks, Alecto outperforms IPCP by \mm{5.47\%}, DOL by \mm{5.65\%}, Bandit3 by \mm{3.67\%}, Bandit6 by \mm{2.32\%}. \mm{Thus, on average, Alecto outperforms the previous state-of-the-art prefetcher selection algorithm---Bandit by 2.76\%. For memory-intensive benchmarks, Alecto outperforms Bandit6 by 5.25\%. These improvements are attributed to the fact that, for most benchmarks, Alecto outperforms other prefetchers or prefetcher selection algorithms. \mmreb{There are a few exceptions, such as mcf and omnetpp, which benefit from PMP's aggressive prefetching instructed by Bandit. Alecto falls below Bandit in these cases because it employs a moderate approach that is not overly aggressive. To validate this, we conducted experiments where we lowered the DB for PMP and fixed PMP's prefetching degree in Alecto to 6, the same as Bandit6. The results show that Alecto can achieve similar performance gains (permance gap $<$ 1\%) to Bandit6 for mcf and omnetpp, indicating that we can leverage Control and Status Register (CSR) to fine-tune Alecto’s behavior on specific workloads.}}

To further validate Alecto's efficiency, we assess the key performance metrics of prefetchers across the entire suite of SPEC benchmarks. Figure~\ref{fig:metric} displays the distribution of covered misses with timely prefetches, covered misses with untimely prefetches, uncovered misses, and overpredicted prefetches for all the prefetcher selection algorithms under review. These metrics provide insights into the prefetching accuracy, coverage, and timeliness of each algorithm. 


The experimental results indicate that \textbf{Alecto is more adept at harmonizing prefetching accuracy, coverage, and timeliness, showcasing superior balance over competing \mm{prefetchers} and prefetcher selection algorithms.} Alecto obtains notable advancements in prefetching accuracy, \mm{surpassing Bandit6 by 13.51\%.} These enhancements are achieved without compromising prefetching coverage and timeliness.

\subsection{Generality: Alecto on Diverse Prefetchers}
\label{subsec:gen2}

\begin{figure}[!t]
\vspace{.1in}
\centering
\includegraphics[width=.99\linewidth]{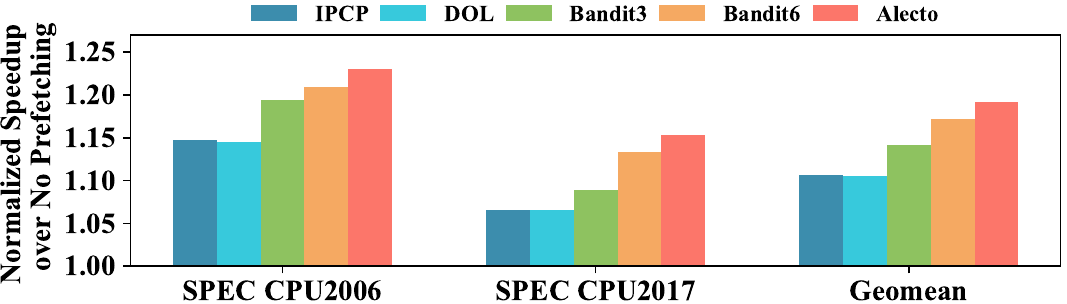}
\vspace{-.25in}
\caption{\mm{Performance speedup when using prefetcher selection algorithms schedule another composite prefetcher: GS \cite{pakalapati2020bouquet} + Berti \cite{navarro2022berti} + CPLX \cite{pakalapati2020bouquet}.}}
\vspace{-.25in}
\label{fig:gen}
\end{figure}

To validate Alecto's ability to schedule various prefetchers, we conduct experiments where the original stride prefetcher (CS) and spatial prefetcher (PMP) are substituted with Berti \cite{navarro2022berti} and CPLX \cite{pakalapati2020bouquet}. PMP, Berti, and CPLX, originating from SMS \cite{somogyi2006spatial}, BOP \cite{michaud2016best}, and VLDP \cite{shevgoor2015efficiently} respectively, represent three state-of-the-art approaches in spatial prefetching. Figure~\ref{fig:gen} shows the performance speedup achieved with various prefetcher selection algorithms using this new composite prefetcher. On average, Alecto outperforms IPCP by 8.52\%, DOL by 8.68\%, Bandit3 by 5.02\%, and Bandit6 by 2.04\%. 

After analyzing the behaviors of each spatial prefetcher under different selection algorithms, we observe PMP and CPLX show superior performance with Alecto compared to Bandit. In contrast, Berti performs well under both Alecto and Bandit, reducing the performance gap between them. Given the more aggressive nature of PMP and CPLX, they require precise selection strategies to mitigate the risk of cache pollution. Berti, known for its accuracy and less aggressive prefetching behavior, is less likely to cause cache pollution. However, this does not imply that aggressive prefetching should be avoided. Instead, when combined with meticulous prefetcher selection, aggressive prefetching can substantially enhance prefetching accuracy, coverage, and timeliness, often outperforming conservative approaches.

\subsection{\mmreb{Effectiveness: Comparison with Non-composite Prefetchers}}

\begin{figure}[h]
\vspace{-.1in}
\centering
\includegraphics[width=.99\linewidth]{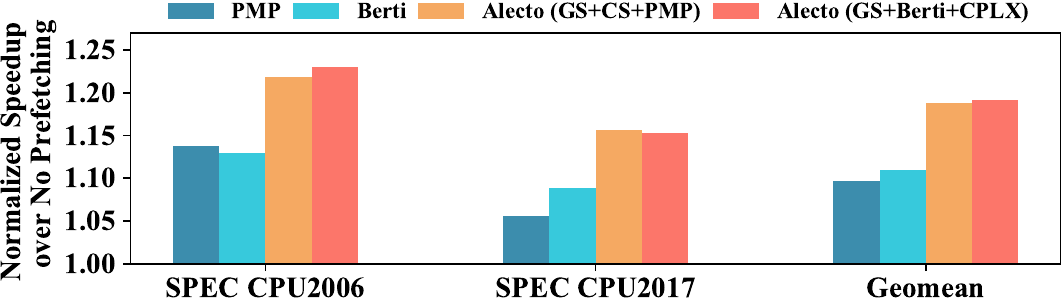}
\vspace{-.25in}
\caption{\mmreb{Performance speedup with non-composite prefetchers.}}
\vspace{-.1in}
\label{fig:individual}
\end{figure}

\mmreb{Figure~\ref{fig:individual} compare two composite prefetchers scheduled by Alecto (evaluated in Section~\ref{subsec:single} and Section~\ref{subsec:gen2}) with the state-of-the-art non-composite prefetchers---PMP and Berti. Alecto (GS+CS+PMP) outperforms PMP by 9.10\% and Berti by 7.83\%. Alecto (GS+Berti+CPLX) outperforms PMP by 9.53\% and Berti by 8.26\%. We observe that composite prefetchers show superior performance to non-composite prefetchers. Therefore, our results indicate the potential of composite prefetchers and highlight the value of relevant research. }

\subsection{Generality: Alecto on Temporal Prefetching}
\label{subsec:gen}

\begin{figure}[!b]
\vspace{-.25in}
\centering
\includegraphics[width=.99\linewidth]{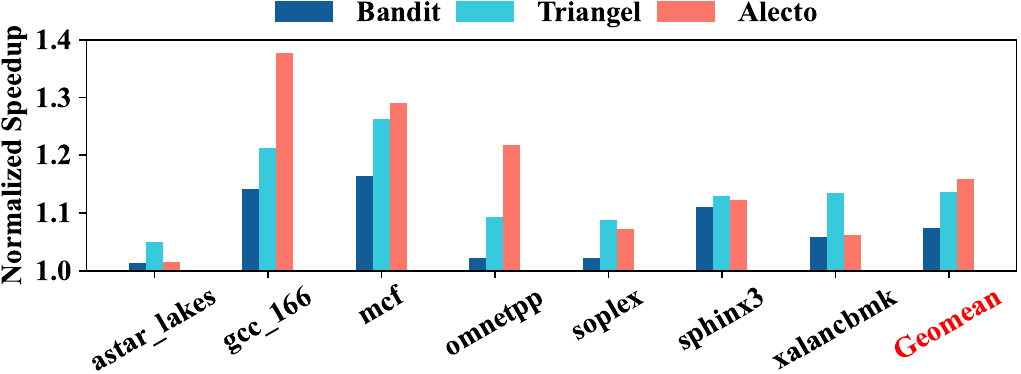}
\vspace{-.25in}
\caption{\mm{The performance speedup of temporal prefetching with different demand request allocation policies.}}
\label{fig:tp_re}
\end{figure}

\begin{figure}[!t]
\vspace{.1in}
\centering
\includegraphics[width=.99\linewidth]{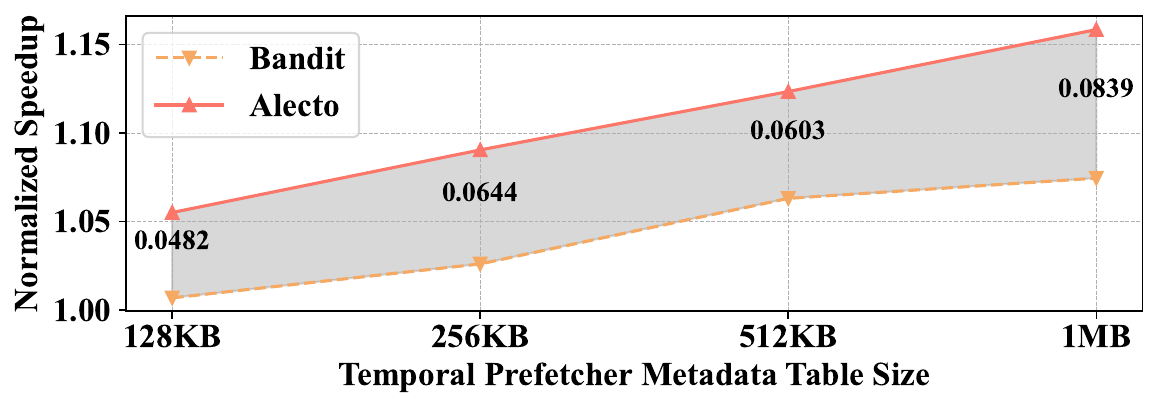}
\vspace{-.25in}
\caption{\mmreb{Geomean IPC speedup with varying metadata table size.}}
\vspace{-.25in}
\label{fig:meta_sen}
\end{figure}

We use the configurations shown in Figure~\ref{fig:tp_exp} to demonstrate the efficiency of Alecto in managing temporal prefetching. Figure~\ref{fig:tp_re} displays the performance speedup, calculated by dividing IPC when both L1 composite prefetchers and an L2 temporal prefetcher are enabled, by the IPC obtained when only L1 composite prefetchers are enabled. Following the methodology of previous studies \cite{wu2019efficient,wu2019temporal,wu2021practical}, our experiments are conducted on representative benchmarks that exhibit temporal patterns. Our results show that Alecto outperforms Bandit by 8.39\%, Triangel by 2.18\%. 

Overall, Alecto's dynamic demand request allocation technique greatly enhances the utilization of metadata storage. Given the long latency of accessing the metadata storage and limited memory bandwidth, simply adjusting the prefetching degree like Bandit is inefficient for temporal prefetching. Alecto and Triangel both support allocating demand requests to temporal prefetchers. However, Triangel does not filter demand requests that can be handled by non-temporal prefetchers (Section~\ref{subsec:temporal}). This results in its inefficiency. 

\mmreb{To highlight the importance of demand request allocation for temporal prefetching, we compare Alecto to Bandit across varying metadata table sizes. As described in Section~\ref{subsec:tp_config}, we disable the resizing of the metadata table for Alecto and Bandit. We vary the metadata table size while maintaining a fixed LLC size. Figure~\ref{fig:meta_sen} demonstrates that Alecto consistently outperforms Bandit under the same metadata budget, with performance gains ranging from 4.82\% to 8.39\%. Furthermore, to achieve the same performance as Bandit with a 1MB metadata table, Alecto only requires less than 256KB.}

\subsection{Sensitivity: LLC Size}
\label{subsec:llcsize}

\begin{figure}[b]
\vspace{-.2in}
\centering
\includegraphics[width=.99\linewidth]{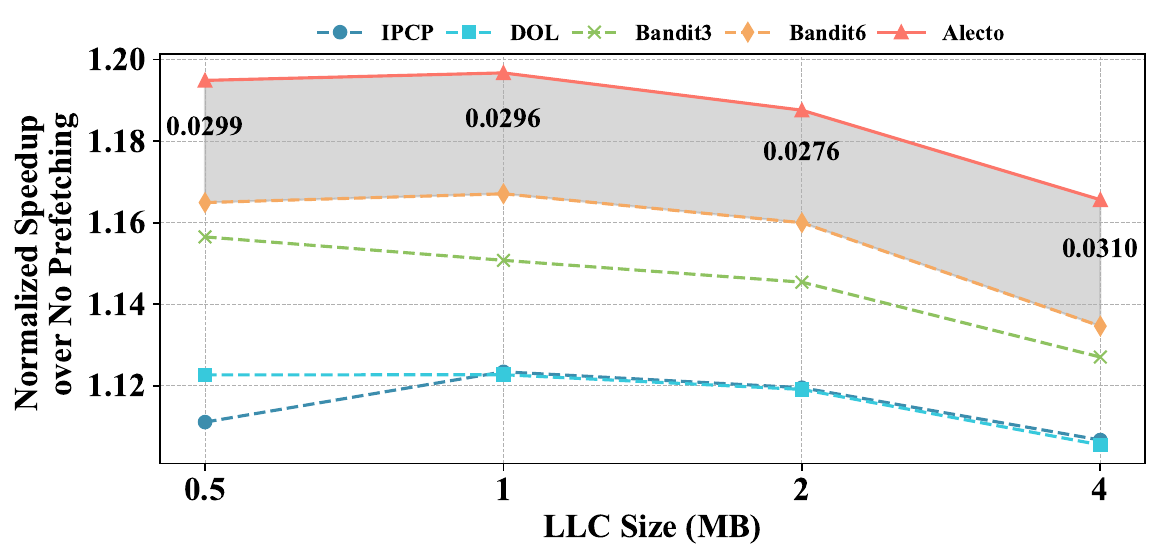}
\vspace{-.3in}
\caption{\mm{Geomean IPC speedup with varying LLC size.}}
\label{fig:llc_sen}
\end{figure}

Figure~\ref{fig:llc_sen} shows the geometric mean of IPC speedup over no prefetching, across various LLC sizes per core for IPCP, DOL, Bandit3, Bandit6, and Alecto. While an increase in LLC size tends to diminish the benefits provided by prefetching, Alecto consistently maintains superior performance over other prefetcher selection algorithms, showcasing its effectiveness regardless of the LLC size. The performance gain of Alecto over Bandit6 varies between \mm{2.76\%} and \mm{3.10\%}. \mm{Notably, Alecto's performance advantage over Bandit does not decrease as the LLC size grows. This suggests that Alecto tends to provide high prefetching accuracy with smaller LLCs, and conversely, it can dynamically increase its aggressiveness in response to low cache pollution levels.}

\subsection{Sensitivity: Memory Bandwidth}
\label{subsec:band}

\begin{figure}[t]
\centering
\includegraphics[width=\linewidth]{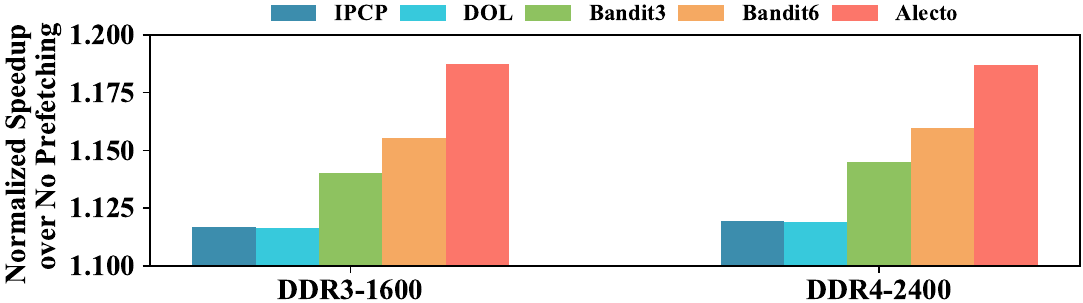}
\vspace{-.25in}
\caption{\mm{Geomean IPC speedup with varying DRAM bandwidth.}}
\vspace{-.25in}
\label{fig:ddr_sen}
\end{figure}

Figure~\ref{fig:ddr_sen} displays the geometric mean of IPC speedup over scenarios without prefetching, comparing two DRAM bandwidth configurations for IPCP, DOL, Bandit3, Bandit6, and Alecto. We have two observations. First, Alecto outperforms other algorithms in both of these configurations. Specifically, Alecto surpasses Bandit6 by \mm{3.18\%} in DDR3-1600, and \mm{2.76\%} in DDR4-2400. Second, Alecto and Bandit6 can benefit more from the elevated memory bandwidth than other prefetcher selection algorithms. For Bandit6, the primary reason for its performance improvement is attributed to its more aggressive prefetching strategy. In contrast, Alecto's performance stems from its ability to balance prefetching accuracy and timeliness. 

\subsection{Scalability: Multi-core Evaluation}
\label{subsec:multi}

\begin{figure}[b]
\vspace{-.15in}
\centering
\includegraphics[width=\linewidth]{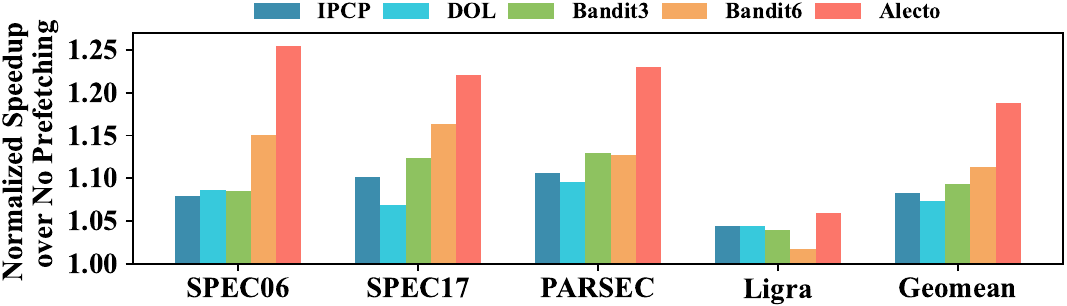}
\vspace{-.25in}
\caption{\mmreb{Eight-core performance speedup compared to no prefetching on SPEC06, SPEC17, PARSEC and Ligra benchmarks.}}
\label{fig:mul}
\end{figure}


Figure~\ref{fig:mul} shows the performance speedup across all prefetcher selection algorithms on an eight-core setup. \mmreb{On average, Alector outperforms IPCP by 10.60\%, DOL by 11.52\%, Bandit3 by 9.51\%, and Bandit6 by 7.56\%.} The performance gap between the Alecto and Bandit becomes enlarged in the eight-core scenario. With the increased core count, Bandit suffers from its coarse-grained prefetcher selection. To mitigate the scalability issue common to RL algorithms, Bandit uses sampled IPC as the sole reward for training its nTable and rTable. As noted in \cite{gerogiannis2023micro}, such sampling is prone to interference between different cores. One core's aggressive prefetching would make other cores get trapped in suboptimal prefetching strategies. In contrast, Alecto controls each prefetcher's prefetching degree in a fine-grained manner, effectively balancing the overall prefetching performance.

\subsection{Scalability: Storage Overhead}
\label{subsec:storage}


\begin{table}[!h]
\centering
\vspace{-.1in}
\caption{Storage overhead analysis.}
\vspace{-.1in}
\label{tab:storage}
\small
\resizebox{0.49\textwidth}{!}{
\begin{tabular}{|c|c|c|c|}
\hline
\textbf{Structure} & \textbf{Entry} & \textbf{Component} & \textbf{Storage} \\
\hline
\hline 
\multirow{3}{*}{\makecell{Allocation Table}} & \multirow{3}{*}{64} & Valid (1 bit) & \multirow{3}{*}{\makecell{$ 640 + 256 \times P $}}  \\
&& Tag (9 bits) & \\
&& States ($4 \times P$ bits) & \\
\hline
\multirow{6}{*}{Sample Table} & \multirow{6}{*}{64} & Valid (1 bit) & \multirow{6}{*}{$1600 + 1024 \times P $} \\
&& Tag (9 bits) & \\
&& Issued ($8 \times P$ bits) & \\
&& Confirmed ($8 \times P$ bits) & \\
&& Deads (7 bits) & \\
&& Demands (8 bits) & \\
\hline
Sandbox Table & \multirow{2}{*}{512} & Tag (6 bits) & \multirow{2}{*}{$ 3072 + 512 \times P $} \\ 
(Prefetch Filter) && Valid for $P_{i}$ (P bits) & \\ 
\hline
\multicolumn{4}{|l|}{Overall: $ 5312 + 1792 \times P$ bits ($\approx $ 1.30~KB when P = 3)} \\
\multicolumn{4}{|l|}{Exclude Sandbox Table: $ 2240 + 1280 \times P$ bits ($\approx$ 760~B when P = 3)} \\
\hline
\end{tabular}
}
\vspace{-.25in}
\end{table}

Table~\ref{tab:storage} details the storage requirements for each component within Alecto. The total storage overhead is 5312 + $1792 \times P$ bits, where $P$ represents the number of prefetchers, demonstrating a linear scalability with the number of prefetchers. Notably, the Sandbox Table serves a dual purpose by also acting as a prefetch filter, a feature that is essential in all cache prefetcher systems, thereby offsetting the need for additional storage to accommodate a separate prefetch filter. Excluding the Sandbox Table, Alecto necessitates only $2240 + 1280 \times P$ bits for the Allocation Table and Sample Table. For the configurations used in our experiments (P=3), the total storage overhead is approximately 760 bytes. 

\mm{Alecto is more scalable than RL-based prefetcher selection algorithms, which typically face exponential growth in storage requirements as the number of selected prefetchers increases. Even Bandit \cite{gerogiannis2023micro}, a relatively lightweight RL-based prefetcher selection algorithm, is not exempt from scalability challenges. Its storage overhead is calculated as $8 \times \#arm$ bytes, where $\#arm = \#actions^{\#prefetcher}$. For Bandit, the actions correspond to the adjusted prefetching degrees. In Alecto, the prefetching degree of a prefetcher can take on $M + 3$ potential values, such as 0, c, $c + 1$, ..., $c + M + 1$. Extending Bandit's model to include $M + 3$ possible values for the prefetching degree would result in a storage overhead of $8 \times (M + 3)^{\#prefetcher}$. Using the settings from our experiment ($M = 5$), this calculation yields $8 \times 8^{3}$ bytes (4~KB), which is 5.4 times more than Alecto's storage requirements. We have implemented this extended version of Bandit and compared it with Bandit6 and Alecto. We find that its performance is lower than Bandit6 by 0.83\% and Alecto by 3.59\%. Our experimental results suggest that Bandit struggles to converge when too many actions are considered.}


\subsection{Efficiency: Energy Overhead}
\label{subsec:energy}

\begin{figure}[h]
\vspace{-.1in}
\centering
\includegraphics[width=.99\linewidth]{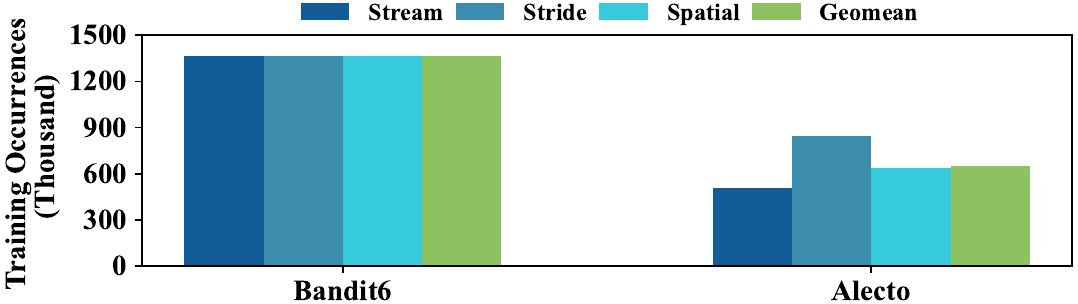}
\vspace{-.25in}
\caption{\mm{The number of each prefetcher's training occurrences in Alecto compared to Bandit.}}
\label{fig:energy}
\end{figure}

We evaluate the energy efficiency of Alecto at both the system and prefetcher levels. At the system level, we utilize CACTI~\cite{muralimanohar2009cacti} to model the energy consumption of the memory hierarchy under a $\qty{22}{nm}$ technology node. At the prefetcher level, we estimate the energy consumption of prefetchers based on their training occurrences. This estimation is supported by two key observations: First, the dynamic power of prefetchers accounts for the vast majority of their total power consumption. Second, prefetchers' primary source of dynamic power consumption is accessing their internal tables, which directly correlates with their training occurrences. \qjreb{Our experiments show that, at the system level, Alecto achieves an average energy reduction of 7\% for the memory hierarchy compared to Bandit6. At the prefetcher level, as shown in Figure~\ref{fig:energy}, Alecto reduces training occurrences of each prefetcher by 48\%. }


\section{\mmreb{Discussion}}
\label{sec:discussion}

\subsection{\mmreb{Alecto's Features: Ablation Study}}
\label{subsec:ablation}

\begin{figure}[!t]
\vspace{-.1in}
\centering
\includegraphics[width=.99\linewidth]{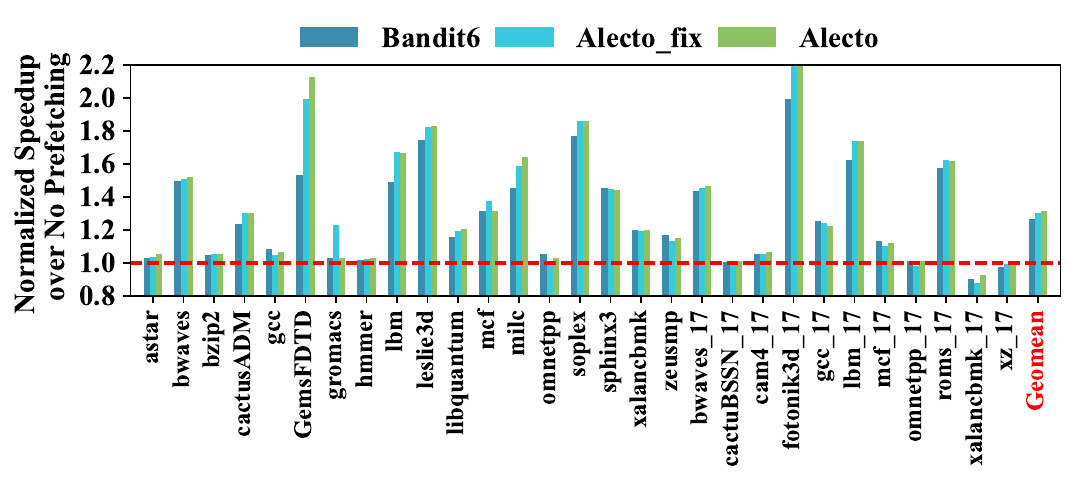}
\vspace{-.25in}
\caption{\mmreb{On memory intensive benchmarks, performance speedup for Bandit6, Alecto with the fixed prefetching degree, and complete Alecto.}}
\vspace{-.25in}
\label{fig:fix}
\end{figure}

\mmreb{We prove that the primary performance gain of Alecto is contributed by demand request allocation. Alecto consists of two parts: (1) demand request allocation; and (2) dynamic prefetching degree adjustment for the allocated prefetchers. To study the impact of each component on overall performance, we modify the state machine in Figure~\ref{fig:state_machine} to isolate Alecto with prefetching degree adjustment. Specifically, we decouple the IA\_m state with the prefetchers' prefetching degree. Once a prefetcher is promoted to IA\_m state, it is configured to issue up to 6 prefetch requests, similar to Bandit6. Figure~\ref{fig:fix} shows that Alecto without prefetching degree adjustment outperforms Bandit6 by 4.34\%, compared to a 5.25\% when prefetching degree adjustment is enabled (Section~\ref{subsec:single}).}

\subsection{\mmreb{Prefetching Degree Study}}

\mmreb{To further validate the results discussed in Section~\ref{subsec:ablation}, we calculate the average prefetching degree for Alecto and compare it to Bandit6. The stream prefetcher scheduled by Alecto issues 79\% of the prefetch requests, the stride prefetcher issues 124\%, and the spatial prefetcher issues 94\%. This indicates Alecto's overall aggressiveness is comparable to Bandit6. Additionally, the temporal prefetcher issues 156\%, demonstrating that the temporal prefetcher under Alecto receives better training than Bandit6, because (1) the temporal prefetcher is limited to issue a maximum of one prefetch request per training occurrence (Section~\ref{subsec:tp_config}), and (2) the temporal prefetcher under Alecto thus have more opportunities to generate prefetch requests.}

\subsection{\mmreb{Prefetch Filtering Study}}

\begin{figure}[!t]
\vspace{-.1in}
\centering
\includegraphics[width=.99\linewidth]{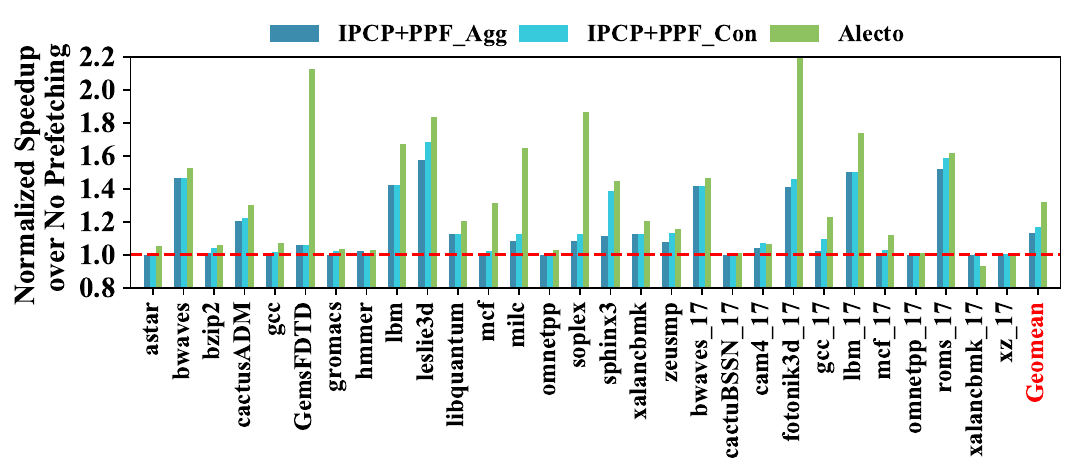}
\vspace{-.25in}
\caption{\mmreb{On memory intensive benchmarks, performance speedup for IPCP, IPCP+PPF\_Aggressive, IPCP+PPF\_Conservative, and Alecto. They schedule the same composite prefetcher: GS+CS+PMP.}}
\vspace{-.25in}
\label{fig:ppf}
\end{figure}

\mmreb{We prove that dynamic demand request allocation is more effective than prefetch filtering only, such as PPF (Perceptron-based Prefetch Filtering) \cite{bhatia2019perceptron}. Alecto's dynamic demand request allocation allows each scheduled prefetcher to receive better training while simultaneously filtering out unnecessary prefetch requests that would otherwise be generated during training. Figure~\ref{fig:ppf} compares Alecto to PPF \cite{bhatia2019perceptron}. We use features from PPF's original paper \cite{bhatia2019perceptron} to train PPF. We apply PPF to filter requests generated by the composite prefetcher. To isolate the impact of prefetcher selection, we use IPCP to schedule the composite prefetcher in PPF's configuration. }

\mmreb{We tune PPF into two versions: PPF\_Aggressive and PPF\_Conservative. The former is more aggressive in identifying and filtering useless prefetch requests than the latter. Our experiments reveal that though PPF improves prefetching accuracy, it incorrectly filters out many useful prefetch requests. For example, in the case of PPF\_Aggressive on GemsFDTD, prefetching accuracy increases from 0.53 to 0.9, but prefetching coverage drops from 0.67 to 0.35. Consequently, Alecto outperforms IPCP+PPF\_Aggressive by 18.38\%, and IPCP+PPF\_Conservative by 14.98\% across all memory intensive workloads. Alecto's dynamic demand request allocation increases both the prefetching accuracy and coverage (Section~\ref{subsec:single}), resulting in better performance.}

\section{Conclusion}
\label{sec:conclusion}
\mmreb{In this paper, we proposed Alecto, a prefetcher selection framework with dynamic demand request allocation. Alecto adopts fine-grained prefetchers identification that identifies suitable and unsuitable prefetchers for each memory access instruction.  Based on the identification, Alecto can tailor optimal prefetching strategies to each memory access instruction, thereby optimizing the overall prefetching accuracy, coverage, and timeliness. We quantified the performance of Alecto across extensive prefetcher settings and benchmarks, which showed that Alecto outperforms state-of-the-art prefetcher selection algorithms and non-composite individual prefetchers.}

\section*{Acknowledgements}
This work is supported by Hong Kong Research Grants Council (RGC) ECS Grant 26208723, National Natural Science Foundation of China (92364102, 62304192), and ACCESS – AI Chip Center for Emerging Smart Systems, sponsored by InnoHK funding, Hong Kong SAR.

\bibliographystyle{IEEEtranS}
\bibliography{refs}


\end{document}